\documentclass[aps,prl,twocolumn,groupedaddress,showpacs]{revtex4}

\usepackage{dcolumn}
\usepackage{amsmath}
\usepackage{bm} 
\usepackage{graphicx}  
\usepackage{float}

\begin{document} 


\title{Vacuum-Induced Symmetry Breaking in a Superconducting Quantum Circuit}


\author{L. Garziano$^1$, R. Stassi$^1$, A. Ridolfo$^1$,  O. Di Stefano$^1$, S. Savasta$^1$}
\affiliation{$^1$Dipartimento di Fisica e di Scienze della Terra, Universit\`{a} di Messina, Viale F. Stagno d'Alcontres 31, I-98166 Messina, Italy}


\date{\today}

\begin{abstract}  
 The ultrastrong-coupling regime, where the atom-cavity coupling rate reaches a considerable fraction
of the cavity or atom transition frequencies, has been reported in a flux qubit superconducting quantum circuit coupled to an on-chip coplanar resonator.
In this regime situations may arise where the resonator field $\hat X = \hat a+ \hat a^\dag$ acquires a nonzero expectation value in the system ground state.
We demonstrate  that in this case the parity symmetry of an additional artificial atom with an even potential is broken by the interaction with the resonator.
Such mechanism is analogous to the Higgs mechanism where the gauge symmetry of the weak force's gauge bosons is broken by the nonzero vacuum expectation value of the Higgs field.
The results here presented open the way to controllable experiments on symmetry breaking mechanisms induced by nonzero vacuum expectation values. Moreover the here proposed mechanism can be used as a probe of the ground state macroscopic coherence emerging from quantum phase transitions  with vacuum degeneracy.
\end{abstract}

\pacs{ 42.50.Pq, 03.70.+k, 05.30.Rt}

\maketitle

Superconducting circuits based on Josephson junctions received considerable attention during recent years
because they are considered  promising fundamental building blocks
for quantum computing \cite{Makhlin2001,You2006,Clarke2008,Schoelkopf2008}.   
Experimental
progress on superconducting resonator-qubit systems
have also inspired theoretical and
experimental investigations of quantum optics in the microwave
regime \cite{You2011}. These recent advances in the engineering and control of quantum fields in superconducting
circuits have also opened up the possibility to explore quantum vacuum effects with these devices \cite{Norirmp,Fragner2008,Wilson2011,Lahteenmaki2013}.
Superconducting circuits have also been used  for the realization of  systems with ultrastrong
qubit-resonator coupling \cite{Niemczyk2010,Forn-Diaz2010}. This regime, where the coupling rate $g$ becomes comparable to the resonance frequency of the fundamental resonator mode, is attracting interest  because of the possibility of manipulating the cavity quantum electrodynamic vacuum with controllable physical properties and it has been realized in a variety of solid state quantum systems \cite{Anappara2008,Gunter2009,Todorov2010,Schwartz2011,Geiser2012,Scalari2012}.
The most puzzling property of light-matter systems displaying ultrastrong coupling (USC) is that their ground state $| G \rangle$ is a squeezed
vacuum containing correlated pairs of matter excitations and cavity photons \cite{Ashhab2010,Quattropani}. 
The photons in the ground state $| G \rangle$ are, however, virtual and cannot be detected, although the intracavity mean photon number is not zero, $\langle \hat a^\dag \hat a \rangle \neq 0$. However it has been theoretically shown that in the presence of non adiabatic modulations, induced Raman transitions, sudden on/off switches of the light-matter interaction or spontaneous decay mechanisms these virtual photons can be converted to real ones giving rise to a stream of quantum vacuum radiation (see e.g. \cite{DeLiberato2009,Palma2012,Stassi2013,Garziano2013,pracinesi2014}).


Here we investigate the cavity quantum electrodynamic ground state and its properties in  superconducting resonators coupled with one or more flux qubits subject to 
parity symmetry breaking. The symmetry breaking can be induced by an external flux $\Phi_{\rm ext} \neq \Phi_0 /2$ threading the artificial atoms, where  $\Phi_0$ is the magnetic flux quantum \cite{Gross2008,Liu2005}, or can be spontaneous, arising from the degeneracy or the quasidegeneracy of the ground state in USC cavity-QED systems \cite{Brandes2004, NatafPRL, Natafnature}. In the latter case the two degenerate ground states are the product of a ``ferromagnetic" state for the artificial atoms and coherent states for the resonator modes.
In the USC regime the resonator field $\hat X = \hat a+ \hat a^\dag$ acquires a nonzero vacuum expectation value ($\langle G | \hat X | G \rangle \neq 0$) displaying a high degree of  first order coherence $|\langle G | \hat a | G \rangle|^2/ \langle G | \hat a^\dag \hat a | G \rangle \simeq  1$  also when the symmetry breaking is induced by an external flux offset $f =\Phi_{\rm ext} - \Phi_0/2$.
The coherence properties of this peculiar ground states can be directly probed by fast modulating or switching the resonator-qubit interaction rate.  
Specifically we show that such a modulation or switch gives rise to quantum vacuum radiation with a high degree of first-order optical coherence. The strong coupling between states with different number of excitations determined by symmetry breaking  induced by a flux offset in the USC regime has been demonstrated in a flux qubit interacting with an on-chip resonator in the USC regime \cite{Niemczyk2010}.

The nonzero vacuum expectation value of the resonator field is reminescent of the Higgs scalar field $\hat \phi$
whose expectation value in the vacuum state $\langle 0 | \hat \phi | 0 \rangle = v$ is different from zero \cite{Higgs}. 
In the standard model, the $W^\pm$ and $Z$ weak gauge bosons would  be massless as a consequence of the electroweak ($SU(2)\times U(1)$) gauge symmetry.
However, according to the Higgs mechanism, the gauge symmetry  is broken and the mass is gained by the interaction of gauge bosons with a scalar field (so-called Higgs field) having a nonzero vacuum expectation value as a consequence of spontaneous symmetry breaking \cite{Weinberg1967}.
Here we show that the parity symmetry of an artificial atom with an even potential is broken by the interaction with a resonator with a nonzero field  vacuum expectation value $\langle G | \hat X | G \rangle \neq 0$ in close analogy with the Higgs mechanism (Fig.1). The signature of symmetry breaking is the appearance of two-photon transitions among levels with opposite parity which are forbidden by parity selection rules (See Appendix A). 
It is worth noting that of course it is not possible to reproduce the Higgs model with just a single mode resonator and a few two level systems. The Higgs model has more degrees of freedom and distinguishes the Goldstone and Higgs mode, which is not present here. Moreover the gauge theory describes  quantum field theories with Lagrangians invariant under a continuous group of local transformations.  Parity is instead a discrete symmetry. Hence the analogy among the Higgs mechanism and our study consists only of one key feature: in both cases symmetry is broken by a nonzero vacuum (ground state) expectation value of a quantum field.

Spontaneous symmetry breaking is a widespread phenomenon in physics, nevertheless controllable experiments where the nonzero vacuum expectation value of a field breaks the symmetry of a quantum system are absent at our knowledge.
Spontaneous symmetry breaking is an important feature of quantum phase transitions and typically implies the
appearance of degenerate ground states. One example is given by the Dicke model \cite{Brandes2004}, describing a collection
of two-level systems interacting with a single-mode bosonic field. Recently an experimental realization with pump-dressed atoms embedded in an optical cavity has been reported \cite{Baumann2010}. In order to have a Dicke phase transition with a time-independent Hamiltonian and a true ground state, settings based on superconducting circuit QED have been proposed \cite{NatafPRL, Natafnature}.
One drawback however of systems with  these features is that the macroscopic coherence of the resonator field in the ground state cannot give rise to a detectable optical signal \cite{DeLiberato2009,Ridolfo2012}.  
The here  proposed symmetry-breaking mechanism can thus be exploited as a means to probe the occurrence of vacuum coherence in this kind of quantum phase transitions. 

We start considering a quantum circuit constituted by a coplanar resonator interacting with a number of flux qubits in the USC regime (Fig.1). For suitable junction sizes, the qubit potential landscape can be reduced to a double-well potential, where the two minima correspond to
states with clockwise and anticlockwise persistent currents $\pm I_{\rm p}$ \cite{Niemczyk2010}. At $f = 0$, the lowest two energy states are separated by an energy gap $\Delta$. In the qubit eigenbasis, the qubit Hamiltonian reads $\hat H_{\rm q}  =  \hbar \omega_{\rm q} \hat \sigma_z /2 $, where  $\hbar \omega_{\rm q} = \sqrt{\Delta^2 + (2I_{\rm p} f)^2}$ is the qubit transition frequency with $f =\Phi_{\rm ext} - \Phi_0/2$. We note that  the two-level approximation
is well justified because of the large anharmonicity of this superconducting artificial atom.
The fundamental resonator mode is described as a harmonic oscillator $\hat H_{\rm c} = \hbar \omega_{\rm c} (\hat a^\dag \hat a + 1/2)$, where $\omega_{\rm c}$ is the resonance frequency; generalization to include higher frequency modes is straightforward.
\begin{figure}
\includegraphics[height=23 mm]{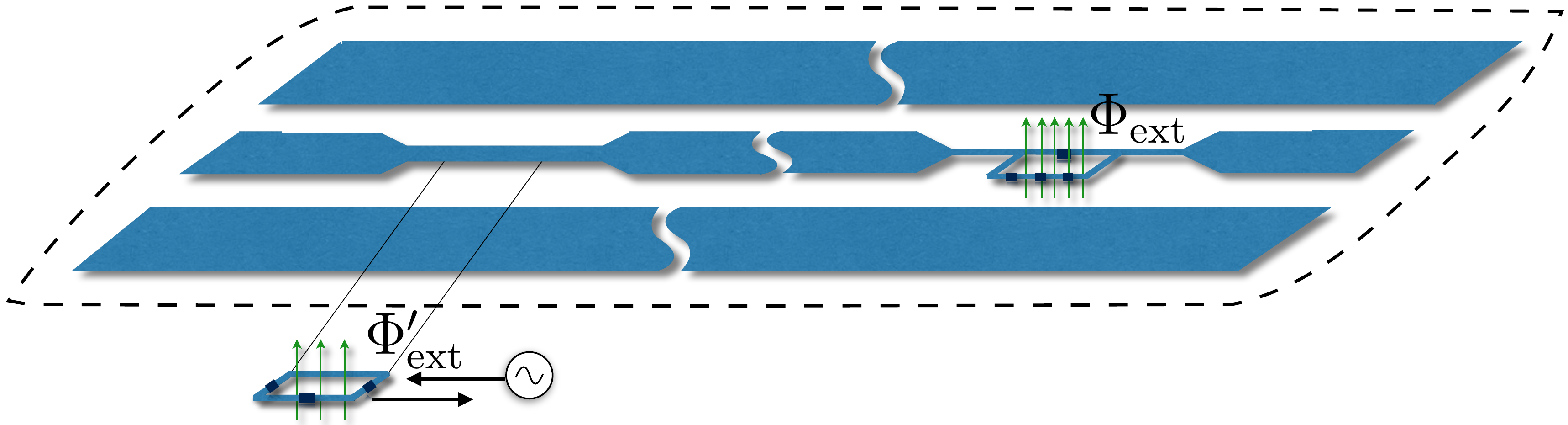}  
\caption{(Color online) Schematic representation of the system. The quantum circuit (dashed black box) is constituted by a coplanar waveguide transmission line resonator ultrastrongly interacting with  embedded identical flux qubits (only one is depicted in the figure).
}   
\label{fig:F1N}  
\end{figure}  

The total Hamiltonian for this quantum circuit can be written as,
\begin{equation}\label{HUSC}
	\hat H_{\rm USC} = \hat H_{\rm c} + \sum_j[\hat H^{(j)}_{\rm q} +  \hbar g_{j} \hat X  (\cos \theta_j \hat \sigma^{(j)}_x + \sin \theta_j \hat \sigma^{(j)}_z )]\, .
\end{equation} 
Here $\hat \sigma^{(j)}_{x,z}$ are Pauli operators for the $j$-th qubit, $g$ is the coupling rate of the qubits
to the cavity mode, and the dependence on the external magnetic fluxes threading the qubits is encoded in the angles $\theta_j$ by the relation $\cos{\theta_j} = \Delta/(\hbar \omega^{(j)}_{\rm q})$. 
When $f=0$ the potentials of the artificial atoms are  symmetric double-well potentials then, below critical coupling rates,  the selection rules are the same as the ones for the electric-dipole transitions in usual atoms. This situation corresponds to $\theta_j = 0$. In this case Eq.\ (\ref{HUSC}) reduces to the standard Dicke Hamiltonian. When $f \neq 0$ ($\theta \neq 0$), the parity symmetry is broken and one- and two- photon transitions can coexist \cite{Gross2008}.
In the following, for the sake of simplicity we will consider equal qubits all with the same coupling rates ($g_j = g$, $\theta_j = \theta$). Qubits displaying different coupling rates do not change the present scenario as shown in Appendix B where approximate results for qubits with different coupling rates are presented. 

The Hamiltonian (\ref{HUSC}) can be diagonalized numerically.  When the coupling rate reaches a considerable fraction of the cavity mode resonance frequency $\omega_{\rm c}$, the ground state $| G \rangle$ acquires a non-negligible amount of excitations. Figure 2a displays a contour plot of the vacuum expectation value $v=\langle G | \hat X | G \rangle$ as a function of the coupling rate $g$ and the angle $\theta$ at positive detuning $\delta =\omega_{\rm q} - \omega_{\rm c} = 0.7\;\omega_{\rm c}$ obtained for a resonator coupled to three qubits. For $\theta = 0$ , the ground state contains only states with an even number of excitations so that $\langle G | \hat X | G \rangle = 0$. On the contrary, when $\theta \neq 0$ the ground state contains also odd excitations so that $v \equiv \langle G | \hat X | G \rangle \neq 0$.
At large coupling rates, when the ground state becomes quasidegenerate \cite{NatafPRL}, the resonator field acquires a significant vacuum expectation value even for $\theta \simeq 0$ when the double well potential of the qubits tends to be even: a clear example of a vacuum field (ground state) expectation value induced by spontaneous symmetry breaking. 
The quasi-degeneracy of the two lowest energy levels implies that even a very low-temperature reservoir could be able to
affect coherence inducing incoherent jumps among the two levels. However it has been shown that specific choices of the system coupling with the reservoir and/or 
reservoir density of states can prevent these transitions \cite{LeggettRMP}.
As mentioned above, these virtual photons in the ground state cannot be detected. 
If it would be possible to abruptly switch-off the coupling rate $g$, the bound (virtual) photons present in the ground state could be released producing an output photon flux $\phi_{\rm out} = \gamma_{\rm c} \langle G | \hat a^\dag \hat a  | G \rangle \geq \gamma_{\rm c} |\langle G | \hat a  | G \rangle|^2$, where $\gamma_{\rm c}$ is the resonator loss rate due to external coupling. 
A more feasible way to detect such an extracavity quantum vacuum radiation is to
apply a harmonic temporal modulation of the cavity-atom coupling rate \cite{DeLiberato2009} of the form $g = g_0 + g_1 \sin {\omega t}$. Such switches and modulations  of the coupling rate the coupling strength $g$, while keeping
the resonance frequency of the flux qubit constant, can be realized by suitable schemes applying external time-dependent magnetic fields \cite{Peropadre2010}.
Figure 2b displays the extracavity emission rate (See Appendix C) from a resonator coupled to a single flux qubit as a function of the modulation frequency $\omega$.
The dashed-dotted line displays
$\phi_{\rm out}/ \gamma_{\rm c}$ for $\theta = \pi /10$. The peaks correspond to the system dressed energy-levels.
The lowest energy peak is a cavity-like peak, while the second one is  the atom-like peak. The dashed curve describes the  coherent part of the emission rate. The lowest energy peak is fully coherent; the Fano-like lineshape of the higher energy peaks origins from interference effects due to the alternate phases ($0$,$\pi$) of the dressed states. 
The results in Figure 2b demonstrate that it is possible to observe quantum vacuum radiation out from vacuum fluctuations with a high degree of first order coherence.
The continuous line describes the emission rate predicted for $\theta = 0$. At such an angle and coupling $g_0$ the coherent part is strictly zero and the first two resonances are not present owing to the parity selection rule. 
\begin{figure}[htpb] 
\includegraphics[height=33 mm]{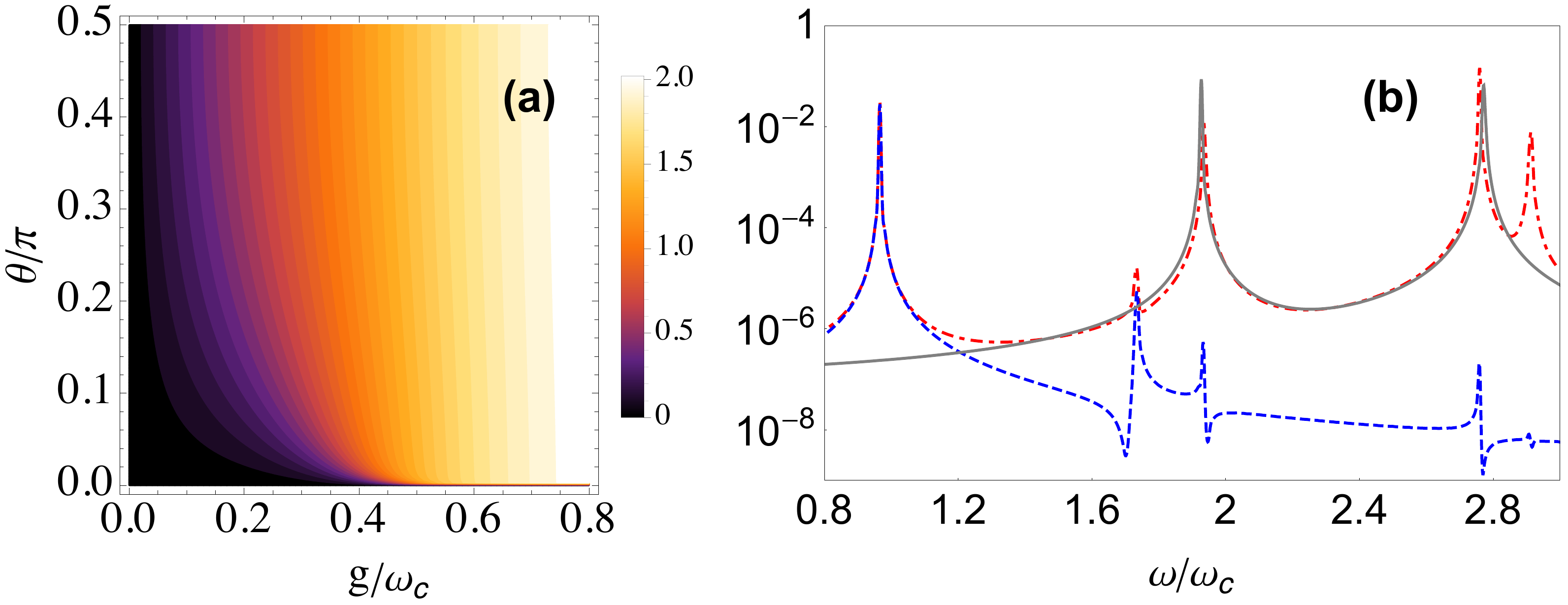}  
\caption{(Color online) a) Contour plot of the vacuum expectation value $v \equiv \langle G | \hat X | G \rangle$ of a resonator- three qubit system in the dispersive regime ($\delta =\omega_{\rm q} - \omega_{\rm c} = 0.7\;\omega_{\rm c}$) as a function of the coupling rate $g$ and the external magnetic flux (encoded in the angle $\theta$) threading the qubit. 
b) Extracavity emission rate $\phi_{\rm out}/ \gamma_{\rm c}$ as a function of the modulation frequency of the resonator-qubit interaction rate $g = g_0 + g_1 \sin {\omega t}$\; for $\theta = 0$ (solid grey line) and $\theta = \pi /10$ (dot-dashed red line). Numerical values for the costant part $g_0$ of the coupling rate and the modulation amplitude $g_1$ are, respectively, $g_0=0.15\; \omega _{\rm c}$ and $g_1=9\times 10^{-4}\;\omega _{\rm c}$. 
The loss rates for the cavity and the qubit are $\gamma_{\rm c}=\gamma_{\rm q}= 10^{-3}\; \omega _{\rm c}$.
The dashed blue line describes the coherent part $|\langle G | \hat a | G \rangle|^2$ of the emission rate for $\theta = \pi /10$. 
} 
\label{fig:F2N}       
\end{figure}    

We now study the influence of the nonzero vacuum expectation value of the resonator on an additional flux qubit (P qubit).
Specifically we consider the P qubit interacting with the resonator at negative detuning $\delta' = \omega'_{\rm q}- \omega_{\rm c} \ll -g'$ (see e.g \cite{Agarwal2013}),
where  $\hbar \omega'_{\rm q} = \Delta'$ is the transition energy of this second qubit and $g' < g$ is its coupling rate with the resonator. This P qubit is subject to an even potential landscape and it can be directly excited by a microwave antenna. The total system
 Hamiltonian can thus be written as
\begin{equation}\label{HUSC1}
	\hat H = \hat  H_{\rm USC} + \Delta' \hat \sigma'_{z}/2 + \hbar g' \hat X\,  \hat \sigma'_{x} + \hbar {\cal E} (t)\,  \hat \sigma'_{x}\, ,
\end{equation}
where $\hat \sigma'_{x(z)}$ are Pauli operators acting on the Hilbert space of the P qubit, the third term in Eq.\ (\ref{HUSC1}) describes the interaction of the qubit with the resonator and the last one describes the qubit excitation by an applied time-dependent microwave electromagnetic field. We start with an approximate  theoretical analysis, then we will show the results of  a full numerical test. In the dispersive regime, if the probe qubit  is excited with a low frequency field  ($\omega << \omega_{\rm c}$), the USC system (resonator plus embedded qubits) can be assumed to be into the ground state. In this case the effective Hamiltonian felt by the probe qubit is $\hat H' = \langle G |\hat H| G\rangle $.  Dropping a constant contribution, $\hat H'$ can be written as $
\hat H'  =  \Delta' \hat \sigma'_{z}/2  + \hbar g' v\,   \hat \sigma'_{x} + \hbar {\cal E} (t)\,  \hat \sigma'_{x}$.
The term $\hbar g' v\,   \hat \sigma'_{x}$ arising from the vacuum expectation value of the resonator field induces a symmetry breaking on the P qubit.  Changing the qubit basis in order to diagonalize the time-independent part of the above equation, we obtain
	$\tilde H'=  \omega'_{\rm q} \hat \sigma'_{z}/2  +\hbar {\cal E}(t) \, 
(\sin{\alpha}\,  \hat \sigma'_z  + \cos \alpha\, \hat \sigma'_{x})$,
where $\omega'_{\rm q} = \sqrt{{\Delta'/ \hbar}^2+ (2g' v)^2}$ and $\sin \alpha =2g' v/ \omega'_{\rm q}$.
By considering a monochromatic excitation signal ${\cal E}(t) = {\cal E}_0 \sin (\omega t)$ and applying standard time-dependent second-order perturbation theory, we obtain the two-photon transition rate
${\cal P}_{g' \to e'} = \pi {\cal E}^2_0\, (\sin^2 2 \alpha)/(\hbar^2 \omega'^2_{\rm q})\delta(2 \omega - \omega'_{\rm q})$.
As expected by dipole selection rules, this transition rate is zero for a zero vacuum expectation value of the resonator field ($v = 0 \Rightarrow \alpha = 0$). On the contrary a nonzero vacuum expectation value activates two-photon transitions giving rise to an absorption peak at $\omega = \omega'_{\rm q}/2$ which is a direct evidence of the parity-symmetry breaking mechanism induced by the vacuum field of the resonator. The two-photon transition rate is maximum at $\alpha = \pi/4$.
As shown in Fig.\ 2a,  the vacuum expectation value $v$ can reach values of the order of 2 already with only three qubits coupled to the resonator.  In the presence of quite large $v$ values, the angle $\alpha$ can reach  nonnegligible values even for a relatively small coupling $g'$ (e.g. $g' = 0.05 \omega_{\rm c}$ and $v =4$ imply $\sin \alpha = 0.4$). 

In order to highlight the analogy with the Higgs mechanism,  let us describe this quantum circuit analogy from the point of view of  experimenters with limited resources. Specifically we assume that they have direct access only to the P qubit and  that in their laboratory they have only excitation sources with a limited energy range $\hbar \omega < \hbar \omega_{\rm c}$. They know that the artificial atom P under investigation has an even potential. They probe the qubit and find a resonance at  $\omega = \omega'_{\rm q}$ being very surprised to find also an additional resonance
at $\omega = \omega'_{\rm q}/2$. The observed coexistence of one- and two-photon
processes is a signature of  parity symmetry-breaking. Hence they start searching for a mechanism explaining it. One of them conjectured that  this effect can originate  from the interaction of the qubit with a quantum field displaying a nonzero vacuum expectation value (the resonator in the USC regime).
This additional quantum field is expected to be in its ground state since in the investigated energy range no additional resonances are found. Moreover they are aware that if this ghost field actually interacts with the qubit,
it should be possible to excite it probing the qubit with an higher energy source. Finally they get such a source and will find at higher energies a new peak corresponding to the dressed resonance frequency of the resonator. Such a peak will be shown in the subsequent full numerical demonstration.  

To provide full evidence for the above reasoning, we present a  numerical demonstration based on
the solution of a master equation in the dressed-states basis that takes into account the influence of the qubits-resonator coupling in the interaction of each subsystem with the environment \cite{Beaudoin2011,Ridolfo2013, Ridolfo2013a}. We calculate numerically the system dynamics under the effect of Hamiltonian (\ref{HUSC1}) and interacting with zero temperature reservoirs (See Appendix D). According to Eq.\ (\ref{HUSC1}) the P qubit  is excited by an applied monochromatic microwave field ${\cal E}(t) = {\cal E}_0 \sin (\omega t)$.  Figure\ 3 displays the  excited state population of the P qubit. The lowest energy peak corresponds to the dressed resonance frequency $\omega'_{\rm q}$ of the P qubit. 
The inset in Fig.\ 3 shows that an additional absorption peak, corresponding to two-photon absorption, is observed at $\omega = \omega'_{\rm q}/2$. As this absorption peak describes a two-photon process, a larger modulation amplitude than that used for obtaining the main spectrum displayed in Fig.\ 3 has been applied.
The appearence of this two-photon transition is a clear signature of the parity simmetry breaking of the P qubit  induced by its interaction with a resonator displaying a nonzero field vacuum expectation value. 
The basic idea under the search for the Higgs boson is that symmetry-breaking fields, when suitably excited, must bring forth characteristic particles: their excitation quanta.
The peak at higher energy  $\simeq \hbar \omega_{\rm c}$ in Fig. 3 arises from the excitation of the quantum field (the resonator in the USC regime) via its interaction with the P qubit. Such a resonance confirms that the violation of parity symmetry of the P qubit  can be attributed to the vacuum expectation value of a quantum field, the analogous of the Higgs field whose resonance energy is at about $\hbar \omega_{\rm c}$. The used parameters are provided in the figure caption. Higher values of two-photon excitation rate can be obtained for resonators coupled in the USC regime with more artificial atoms, even in the presence of a lower coupling with the P qubit. In this case the very interesting situation of an Higgs-like  symmetry breaking mechanism induced by a resonator field vacuum expectation value arising from spontaneous symmetry breaking can be realized. 
\begin{figure}
\includegraphics[height=40mm]{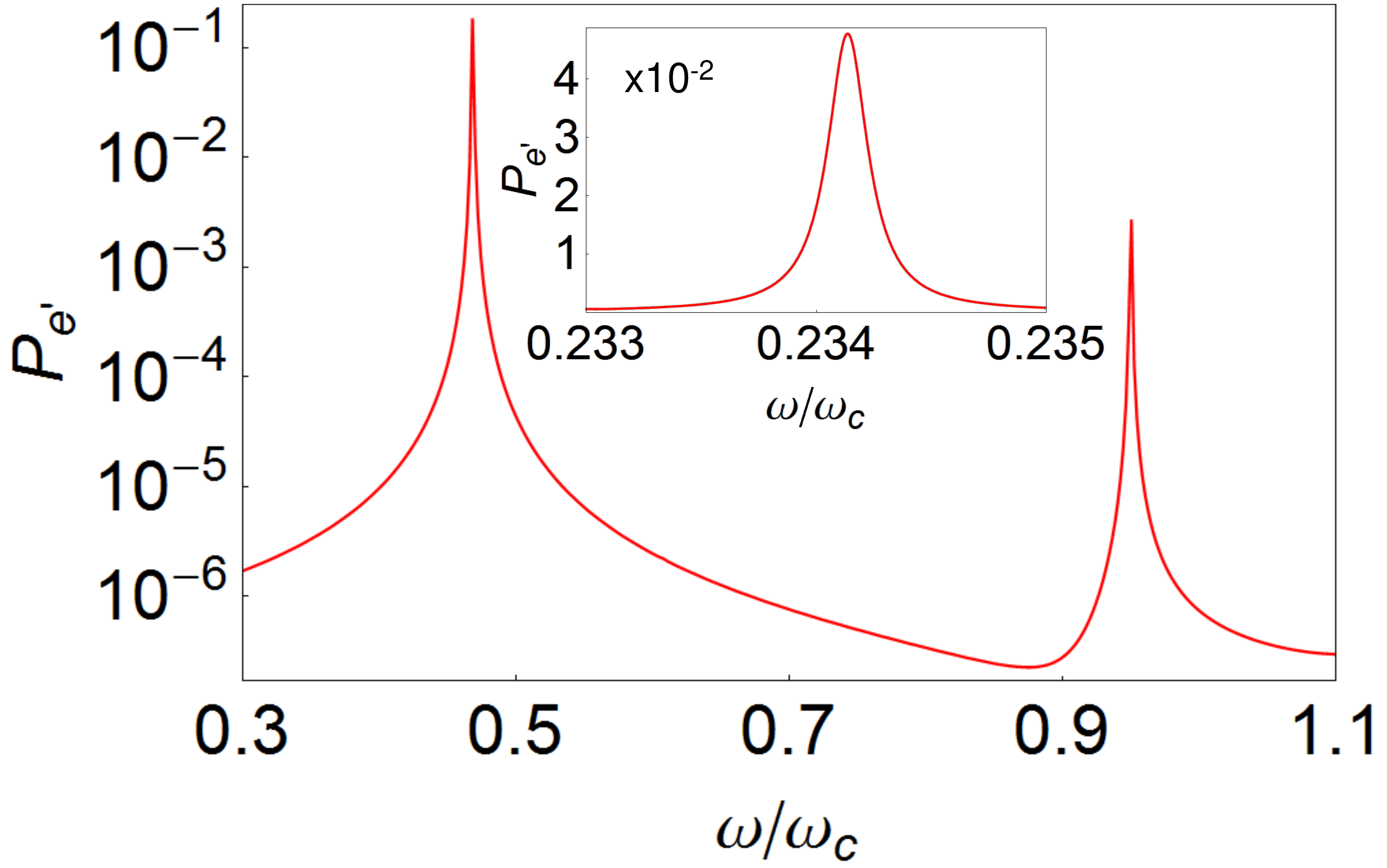}  
\caption{(Color online) Absorption spectrum as a function of the driving frequency of the coherent drive ${\cal E}(t) = {\cal E}_0 \sin (\omega t)$ probing the probe qubit. The resonator is ultrastrongly coupled with one qubit  (at positive detuning $\delta =\omega _{\rm q}-\omega _{\rm c}= 0.7\; \omega _{\rm c}$) and at the P qubit  (at negative detuning $\delta' =\omega _{\rm q'}-\omega _{\rm c}=- 0.6\; \omega _{\rm c}$) with coupling rates $g=0.5\; \omega _{\rm c}$ and $g'=0.2\;\omega _{\rm c}$, respectively. The flux offset applied to the qubit coupled with the resonator corresponds to $\theta =\pi/4$, while it is zero for the P qubit ($\theta =0$). Loss rates for the cavity and the qubits are $\gamma_{\rm c}=\gamma_{\rm q}=\gamma_{\rm q'}=5\times 10^{-4}\; \omega _{\rm c}$. The lowest energy peak corresponds to the dressed resonance frequency $\omega'_{\rm q}$ of the P qubit. The higher energy peak arises from the excitation of the quantum field (the resonator in the USC regime) interacting with the P qubit.}

\label{fig:F3N}   
\end{figure}     

In conclusion we have shown that superconducting circuits can be used to realize a symmetry-breaking quantum vacuum. Specifically we have shown  that the parity symmetry of an  artificial atom (P qubit) with an even potential is broken by the interaction with a resonator field displaying a nonzero vacuum expectation value in close analogy with the symmetry-breaking Higgs mechanism.
We have also shown  that when the P qubit is excited at suitable energy its interaction with the resonator field determines the excitation of a detectable field  excitation, the analog of the so-called Higgs-particle \cite{Higgs,Weinberg1967,Higgs2}.
The present study is based on superconducting quantum circuits, but it can in principle be implemented also with other solid state systems sharing two key features, light matter  ultrastrong coupling and parity symmetry breaking as e.g. asymmetric, doped quantum wells embedded
into a planar photonic cavity \cite{Liberato2013}. The here  proposed Higgs-like mechanism can also  be exploited  to probe the occurrence of vacuum quantum coherence in quantum phase transitions occurring in systems with time independent Hamiltonians and true ground states.   
  
\section{APPENDIX A: RELEVANT ENERGY DRESSED LEVELS AND ELECTRIC DIPOLE TRANSITIONS }

\begin{figure}[h] 
\includegraphics[height=50 mm]{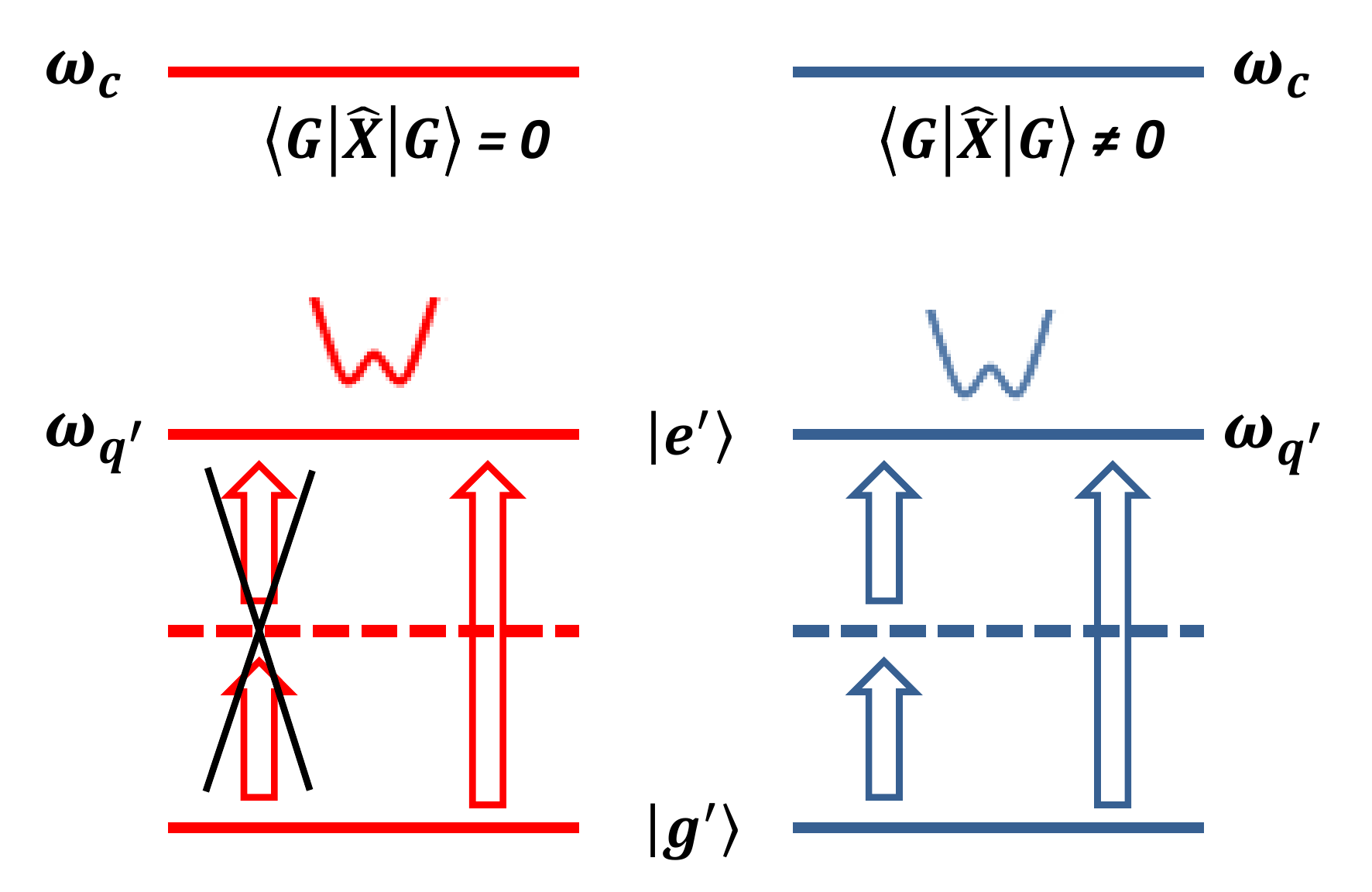}  
\caption{(Color online) Sketch of the relevant dressed energy levels and electric dipole transitions.}  
\label{fig:figlev}  
\end{figure}      

The probe qubit is under the influence of an even potential, as a consequence its quantum states $|{\rm g'} \rangle$ and $| {\rm e'} \rangle$ have definite parity. Hence, according to selection rules for electric-dipole transitions, two photon transitions  $| {\rm g'} \rangle \to |{\rm e'} \rangle$ are forbidden (left panel). However, the parity symmetry of the probe qubit is broken when the probe qubit is coupled to a resonator of higher energy excitations and with  a nonzero vacuum (ground) expectation value $v =\langle {\rm G} |\hat X| {\rm G} \rangle \neq 0$ (right panel) so that two-photon transitions become allowed.

\section{APPENDIX B: ANALYTICAL DERIVATION OF THE VACUUM EXPECTATION VALUE OF THE RESONATOR FIELD}

In this section we consider a system constituted by a coplanar waveguide transmission line resonator ultrastrongly interacting with $N$ embedded flux qubits and derive an analytical expression for the vacuum expectation value of the resonator field $\hat X \equiv \hat a+ \hat a^\dag$.\\ The total system Hamiltonian is given by
\begin{equation}\tag{B1} H_{\rm USC} = \hat H_{\rm c} + \sum_j[\hat H^{(j)}_{\rm q} +  \hbar g_j\; \hat X  (\cos \theta_j  \hat \sigma^{(j)}_x + \sin \theta_j \hat \sigma^{(j)}_z )]\;,
\end{equation}
where $\hat H_{\rm c} = \hbar \omega_{\rm c} (\hat a^\dag \hat a + 1/2)$ describes the cavity mode with resonance frequency $\omega_{\rm c}$ and $\hat H^{(j)}_{\rm q}=\hbar \omega_{\rm q}^{(j)} \hat \sigma^{(j)}_z /2$ is the $j$-th qubit Hamiltonian.
$\omega_{\rm q}^{(j)}$ and \;$\hat \sigma^{(j)}_{x,z}$ are the transition frequency and the Pauli operators for the $j$-th qubit, respectively; $g_j$ is the coupling rate of the $j$-th qubit to the cavity mode, and the dependence on the external magnetic flux threading the $j$-th qubit is encoded in the angle $\theta_j$ by the relation $\cos \theta_j = \Delta/\hbar \omega_{\rm q}^{(j)}$.\\
In the dispersive regime at positive detuning ($\delta_j =\omega_{\rm q}^{(j)} - \omega_{\rm c}>0$), the transition frequencies $\omega_{\rm q}^{(j)}$ are far off resonance and all the qubits can be assumed to be in their ground state ($\sigma^{(j)}_z\cong -1$). This approximation leads to the effective Hamiltonian
\begin{equation}
\label{Heff}\tag{B2}
H^{'}_{\rm USC} \cong  \hat H_{\rm c} - \sum_j  \hbar g_j\;(\hat a+ \hat a^\dag)  \sin \theta_j \;,
\end{equation}
The Hamiltonian (B2) describes a displaced harmonic oscillator and  is easily diagonalized by the following transformation,
\begin{equation}
\label{Trasf}\tag{B3}
\hat{b}=\hat{a}- \sum_j  (g_j\; \sin \theta_j)/ \omega_{\rm c} \;.
\end{equation}
By applying the operator $\hat{b}$ to the ground state $| G \rangle$ of the system we obtain
\begin{equation}\tag{B4}
\hat{b}|G\rangle=(\hat{a}- \sum_j  g_j\; \sin \theta_j/ \omega_{\rm c})|G\rangle=0\, ,
 \end{equation}
so that
\begin{equation}
\label{ground}\tag{B5}
\hat{a}|G\rangle=\sum_j  g_j\; \sin \theta_j/ \omega_{\rm c}|G\rangle\, .
 \end{equation}
Finally, using Eq.(B5) we obtain the approximate analytical expression for the vacuum expectation value of the resonator field
\begin{equation}
\label{vacuum}\tag{B6}
\langle G | \hat X | G \rangle =\langle G | \hat a+ \hat a^\dag | G \rangle= 2\sum_j  g_j\; \sin \theta_j/ \omega_{\rm c}\, .
 \end{equation}

\section{APPENDIX C: PHOTODETECTION IN THE USC REGIME}

One of the most inconvenient issue arising in the ultrastrong coupling regime is that the usual normal order correlation functions fail to describe the output photon emission rate and photon statistics [33]. An incautious application of these standard relations for an USC system in its ground state $|G\rangle$, which now contains a finite number of photons due to the counter-rotating terms in the system Hamiltonian, would therefore predict an unphysical stream of output photons so that $\langle G|\hat a^{\dagger}\hat a|G\rangle\neq 0\ $.

In order to derive the correct output photon emission rate (e.g. physical photons that can be experimentally detected), the cavity-electric field operator $\hat X = \hat a+ \hat a^\dag$ has to be expressed in the atom-cavity dressed basis [34]. Once the field operator is expressed in the energy eigenstates $|j\rangle$\;of the total system Hamiltonian, it can be easily separated in its positive and negative frequency components, $\hat X^{+}$ and $\hat X^{-}$, leading to the relations:

\begin{equation}\tag{C1}\hat X^{+}=\sum_{j,k>j}X_{jk}|j\rangle \langle k| \;; \; \hat X^{-}=(\hat X^{+})^{\dagger} \;,
\end{equation} 
where $X_{jk}\equiv \langle j|\hat a^{\dagger}+\hat a|k\rangle$. 
\\ It is important to notice that the positive frequency component of $\hat X$, according to its actual dynamics, is not simply proportional to the photon annihilation operator $\hat a$. Moreover, the expected result $\hat X^{+}|G\rangle=0$ is correctly obtained for the system in its ground state in contrast to $\hat a|G\rangle\neq 0$.\\ 
According to the input-output theory for resonators, the output field operator can be related through a boundary condition to the intracavity field and the input field operators. 
We consider the specific case of a resonator coupled to a semi-infinite transmission line, pointing out that while the resonator is ultrastrongly coupled with the qubit, its interaction with the reservoir (e.g. the trasmission line) is weak. 
By the Markovian approximation, the input-output relation can be written as
\begin{equation}
\label{in-out}\tag{C2}
	A_{\rm out}(t)=A_{\rm in}(t)-\sqrt{\frac{\hbar \gamma_{\rm c} }{2 \omega_{\rm c}v}} X(t)\, ,
\end{equation} 
where
\begin{equation}
\label{Aout}\tag{C3}
	A_{\rm out}= A_{\rm out}^{+}+A_{\rm out}^{-}\, ,
\end{equation}	
with 	
\begin{equation} 
\label{Aout+}\tag{C4}
	A_{\rm out}^{+}(t)= \frac{1}{2}\int_{0}^{\infty } d\omega \sqrt{\frac{\hbar }{\pi \omega_{\rm c}v}}  a(\omega,t_{1})e^{-i\omega (t-t_{1})}\, 	,
\end{equation}
and $A_{\rm out}^{-}= [A_{\rm out}^{+}]^{\dagger}$.
Here, $v$ is the phase velocity of the trasmission line, the operators $a(\omega,t_{1})$ are the output Bosonic annihilation operators associated to the external (transmission line) continuous modes and $t_1 >t$ can be assumed in the remote future. The rate $\gamma_{\rm c}$ represents the cavity damping due to the interaction of the single cavity mode with the continuum of the external modes. An analogous expression can be written for $A_{\rm in}^{+}(t)$
just replacing in Eq.\ (C4) $t_1$ with $t_0 < t$ in the remote past. 
By taking the Fourier transform of Eq.\ (C2) and summing up only the components at positive frequency, we obtain
\begin{equation}
\label{in-out+}\tag{C5}
	A_{\rm out}^{+}=A_{\rm in}^{+}-\sqrt{\frac{\hbar \gamma_{\rm c} }{2 \omega_{\rm c}v}}X^{+}\, .
\end{equation}
	
By applying the input-output relation (C5), we readily obtain the extracavity emission rate $\langle A_{\rm out}^{-}A_{\rm out}^{+}\rangle $ for the case of an input in the vacuum [23]   
\begin{equation}
\label{rate}\tag{C6}
\frac{2 \omega_{\rm c}v }{\hbar}\langle A_{\rm out}^{-}A_{\rm out}^{+}\rangle= \gamma_{\rm c} \langle X^{-}X^{+}\rangle
\end{equation} \\
Following an analogous procedure, it can be shown that the emission rate from a qubit is proportional to $\langle \hat\sigma_{x}^{'-} \; \hat\sigma_{x}^{'+} \rangle$, where
\begin{equation}\tag{C7}
 \hat\sigma_{x}^{'+}=\sum_{j,k>j}\sigma_{x}^{jk}|j\rangle \langle k| \;; \; \hat\sigma_{x}^{'-}=(\hat\sigma_{x}^{'+})^{\dagger} \;, 
 \end{equation}
with $\sigma_{x}^{jk}\equiv \langle j|\hat\sigma_{x}|k\rangle$.

\section{APPENDIX D: MASTER EQUATION}

In order to correctly describe the quantum dynamics of the system, dissipation induced by its coupling to the environment needs to be considered. Assuming that the cavity and the two level system are weakly coupled with two different baths of harmonic oscillators, the standard approach where the coupling $g$ is ignored while obtaining the dissipative part of the master equation would lead, for a $T=0$ reservoir, to the well known standard master equation. The standard quantum optical master equation can be safely used to describe the dynamics of the system in the weak and strong coupling regimes, in which the ratio $g /\omega _{\rm c}$ is small enough for the RWA approximation to be still valid.\\ 
In the ultrastrong coupling regime however, owing to the high ratio $g /\omega _{\rm c}$, the standard approach fails to correctly describe the dissipation processes and leads to unphysical results as well. In particular, it predicts that even at $T=0$, relaxation would drive the system out of its ground state $|G \rangle$ generating photons in excess to those already present.\\ 
The right procedure that solves such issues consists in taking into account the atom-cavity coupling when deriving the master equation after expressing the Hamiltonian of the system in a basis formed by the eigenstates $|j\rangle$ of the total system Hamiltonian [33,37,38]. The dissipation baths are still treated in the Born-Markov approximation. 
Following this procedure it is possible to obtain the master equation in the dressed picture [37,38]. 
For a $T=0$ reservoir, one obtains:  

\begin{equation} 
\label{4}\tag{D1}
\hat{\dot{\rho }}(t)=-i\left [\hat H_{\rm S},\hat \rho(t) \right ]+\mathcal{L}_{\rm a} \hat \rho(t)+\mathcal{L}_{\rm x}\hat \rho(t)\;.\end{equation}

Here $\mathcal{L}_{\rm a}$ and $\mathcal{L}_{\rm x}$ are the Liouvillian superoperators correctly describing the losses of the system where $\mathcal{L}_{\rm s}\hat \rho(t)=\sum_{j,k>j}\Gamma_{\rm s}^{jk} \mathcal{D}[|j\rangle\langle k|]\hat \rho(t)$ for $s=a,\sigma_{-}$ and $\mathcal{D}[\hat O]\hat\rho=\frac{1}{2}(2 \hat O\hat\rho \hat O^{\dagger}-\hat \rho \hat O^{\dagger}\hat O-\hat O^{\dagger}\hat O \hat \rho)$. $\hat H_{\rm S}$ is, in general, the system Hamiltonian in the absence of the thermal baths. In the limit $g \rightarrow0$, standard dissipators are recovered. 

The relaxation rates $\Gamma_{s}^{jk}=2\pi d_{s}(\Delta_{kj} )\alpha_{s}^{2}(\Delta_{kj} )\left | C_{jk}^{s} \right |^{2}$ depend on the density of states of the baths $d_{\rm s}(\Delta_{kj})$ and the system-bath coupling strength $\alpha_{\rm s}(\Delta_{kj} )$ at the respective transition frequency $\Delta_{kj}\equiv\omega_{k} - \omega_{j}$ as well as on the transition coefficients $C_{jk}=\langle j|\hat s+\hat s^{\dagger}|k\rangle \; (\hat s=\hat a,\hat \sigma_{-} )$. These relaxation coefficients can be interpreted as the full width at half maximum of each $|k\rangle\rightarrow |j\rangle$ transition. In the Born-Markov approximation the density of states of the baths can be considered a slowly varying function of the transition frequencies, so that we can safely assume it to be constant as well as the coupling strength. Specifically, we assume $\alpha_{\rm s}^{2}(\Delta_{kj} )\propto (\Delta_{kj})^{2}$ so that the relaxation coefficients reduce to $\Gamma_{s}^{jk}=\gamma_{s}\left (\frac{\Delta_{kj}}{\omega_{s}}  \right )^{2}C_{jk}^{s}$ where $\gamma_{s}$ are the standard damping rates.  
Equation (D1) can be easily extended to take into account $T\neq0$ reservoirs [38].

\bibliography{VEVprl} 

\end{document}